\documentclass[12pt,preprint]{aastex}
\usepackage{epsf}

\newcommand{\ltsima}{$\buildrel<\over\sim$}
\newcommand{\lapprox}{\lower.5ex\hbox{\ltsima}}
\newcommand{\msun}{M$_{\odot}$}
\newcommand{\kband}{{\it K$_s$}-band}

\shorttitle{Mass Assembly Histories by Morphology in GOODS}
\shortauthors{Bundy et al.}

\begin{document}

\title{The Mass Assembly Histories of Galaxies of Various Morphologies in the GOODS Fields}

\author{Kevin Bundy\altaffilmark{1}, Richard S. Ellis\altaffilmark{1}, Christopher J. Conselice\altaffilmark{1}}

\email{kbundy@astro.caltech.edu, rse@atro.caltech.edu, cc@astro.caltech.edu}

\altaffiltext{1}{105--24 Caltech, 1201 E. California Blvd., Pasadena, CA 91125}


\begin{abstract}
We present an analysis of the growth of stellar mass with cosmic time
partitioned according to galaxy morphology. Using a well-defined catalog
of 2150 galaxies based, in part, on archival data in the Great
Observatories Origins Deep Survey (GOODS) fields, we assign
morphological types in three broad classes (Ellipticals, Spirals,
Peculiar/Irregulars) to a limit of $z_{AB}$=22.5 and make the resulting
catalog publicly available.  Utilizing 893 spectroscopic redshifts,
supplemented by 1013 determined photometrically, we combine optical
photometry from the GOODS catalog and deep \kband~imaging to assign
stellar masses to each galaxy in our sample.  We find little evolution
in the form of the galaxy stellar mass function from $z \sim 1$ to
$z=0$, especially at the high mass end where our results are most
robust.  Although the population of massive galaxies is relatively well
established at $z \sim 1$, its morphological mix continues to change,
with an increasing proportion of early-type galaxies at later times.  By
constructing type-dependent stellar mass functions, we show that in each
of three redshift intervals, E/S0's dominate the higher mass population,
while spirals are favored at lower masses.  This transition occurs at a
stellar mass of 2--3$\times 10^{10}$ \msun~at $z\sim 0.3$ (similar to
local studies) but there is evidence that the relevant mass scale moves
to higher mass at earlier epochs.  Such evolution may represent the
morphological extension of the ``downsizing'' phenomenon, in which the
most massive galaxies stop forming stars first, with lower mass galaxies
becoming quiescent later.  We infer that more massive galaxies evolve
into spheroidal systems at earlier times, and that this morphological
transformation may only be completed 1--2 Gyr after the galaxies emerge
from their active star forming phase.  We discuss several lines of
evidence suggesting that merging may play a key role in generating this
pattern of evolution.

\end{abstract}


\section{Introduction}

Great progress has been made in recent years in defining the star
formation history of galaxies \citep{madau96, blain99}. The
combination of statistically complete redshift surveys \citep{lilly95,
ellis96, steidel99, chapman03} and various diagnostics of star formation
(UV continua, recombination lines and sub-mm emission) has enabled
determinations of the co-moving star formation (SF) density at various
redshifts whose rise and decline around $z\simeq 2$ points to the epoch
when most stars were born \citep[e.g.][]{rudnick03, bouwens03,
bunker04}. Many details remain to be resolved, for example in
reconciling different estimators of star formation
\citep[e.g.][]{sullivan04} and the corrections for dust extinction.  In
addition, recent theoretical work including both numerical simulations
and semi-analytic modeling is in some confusion as to the expected
result \citep[e.g.][]{baugh98, som01, nag04}.

An independent approach to understanding how galaxies form is to conduct
a census of galaxies {\em after} their most active phases and to track
their growing stellar masses. The co-moving stellar mass density at a
given redshift should represent the integral of the previously-discussed
SF density to that epoch, culminating in its locally-determined value
\citep{fuku98}.  Unlike the star formation rate (SFR), the stellar mass
of a galaxy is less transient and can act as a valuable tracer for
evolutionary deductions.

Further insight is gained by tracing the integrated growth in stellar
mass of different populations.  For example, the rapid decline with time
in the SF density over $0<z<1$ \citep{lilly96, fall96} appears to result
from the demise of an abundant population of star forming, irregular
galaxies \citep{glazebrook95, abraham96, B98}. By considering the
declining stellar mass density associated with irregular galaxies,
\citet{BE00} suggested that these sources transform, either by mergers
or other means, into the slowly growing mass identified with regular
ellipticals and spirals.

In a complementary fashion, the recent completion of large infrared
surveys like K20 \citep{cimatti02} and MUNICS \citep{drory01} has traced
the distribution in mass of the most massive galaxies out to $z \sim 2$
\citep{fontana04, drory04}.  These and other studies find a decrease in
the overall normalization of the combined galaxy stellar mass function
with redshift.  \citet{fontana04} find very little evolution in the
shape of the combined mass function out to $z\sim 1$.  \citet{drory04},
in contrast, argue for a stronger evolution out to $z\sim 1$ based on a
decrease in the characteristic mass and steepening of the faint--end
slope.  The MUNICS survey, though extensive in size, relies primarily on
photometric redshifts and, at $K \lesssim 18.7$ (Vega), probes a limited mass
range.  The deeper K20 survey, on the other hand, utilizes primarily
spectroscopic redshifts, but is much smaller and suffers more from
cosmic variance.

With sufficient data, it is possible to combine these earlier
approaches, which concentrated on either volume-integrated quantities of
separate populations or the mass distribution of combined galaxy types,
and construct the stellar mass functions of individual populations.  At
$z=0$, \citet{bell03} have used measurements of concentration and color
in the Sloan Digital Sky Survey (SDSS) to classify galaxies as early or
late type and derived separate mass functions for each, demonstrating
that early-types dominate at higher masses.  \citet{baldry04} use the
bimodal color distribution observed in SDSS to separate early from late
populations and find similar results.  At higher redshifts,
\citet{fontana04} were the first to examine type-dependent mass
functions and divided their sample based on spectral type, finding some
evidence for bimodality.

The public availability of ACS images and photometry in the Great
Observatories Origins Deep Survey \citep[GOODS,][]{gia04} together with
spectroscopic redshifts \citep[e.g.][]{wirth04, lef04} provides a new
opportunity for making progress at $z \sim 1$.  With the addition of
infrared photometry, it is possible to characterize the mass functions
of separate and well defined {\em morphological} populations, although
the small size of the GOODS fields makes cosmic variance one of the
primary sources of uncertainty.  It is therefore advantageous to combine
the two GOODS fields, even though GOODS-N has the benefit of roughly
twice as many spectroscopic redshifts as GOODS-S.  With the various
datasets available in GOODS, we can begin mapping out the mass assembly history
responsible for the origin of the Hubble Sequence as well as
understanding the physical processes that drive this assembly.

A plan of the paper follows. In $\S$\ref{data} we discuss the essential
ingredients: the infrared data, including new \kband~imaging of the
GOODS-N field undertaken with the Hale 5m reflector at Palomar; visual
morphologies of the ACS-selected galaxies in GOODS; and spectroscopic
and photometric redshifts. In $\S$\ref{mass_code} we introduce our
method for measuring stellar masses for galaxies of known redshift based
on infrared and optical photometry.  In $\S$\ref{completeness} we
discuss important issues of completeness and selection effects in the
sample.  In $\S$\ref{results} we present the methods and results of our
analysis of the stellar mass functions and the integrated mass density
of various morphological populations.  We summarize and conclude in
$\S$\ref{conclusions}.  Throughout, we assume a cosmological model with
$\Omega_{\rm M}=0.3$, $\Omega_\Lambda=0.7$ and $H_0=100 h$
km~s$^{-1}$~Mpc$^{-1}$.

\section{Data}\label{data}

This study relies on the combination of many different datasets in the
GOODS fields including infrared observations, spectroscopic and photometric
redshifts, and HST morphologies.  

\subsection{Infrared Imaging}\label{kimaging}

Because deep infrared data is not publicly available in GOODS-N, we
carried out \kband~imaging of the GOODS-N field in three overlapping
pointings 8\farcm6 on a side, using the Wide-field Infrared Camera
\citep[WIRC,][]{wil03} on the Hale 5m telescope at Palomar Observatory.
The observations were made in November 2002, January 2003, and April
2003 under slightly different conditions.  The total integration times
for each of the sub-fields of 15ks, 13ks, and 5.6ks account for the
different observing conditions so that the final depth in each pointing
is similar.  Mosaics of numerous coadditions of a sequence of $4 \times
30$ second exposures were taken in a non-repeating pattern of roughly
7\farcs0 dithers and processed using a double-pass reduction pipeline we
developed specifically for WIRC.  The individual observations were
combined by applying weights based on the seeing, transparency, and
background for each observation.  The final data quality is excellent,
with average FWHM values for stars in the WIRC images of 0\farcs85 for
the first two fields and 1\farcs0 for the third.  Because the WIRC
camera field is fixed North-South and cannot be rotated and only three
positions were imaged, the overlap with the GOODS-N region is only $\sim
70\%$.

The WIRC images were calibrated by observing standard stars during
photometric conditions and checked with comparisons to published K-band
photometry from \citet{fran98}.  A comparison for relatively bright
stars in each field was also made with 2MASS.  Photometric errors and
the image depth were estimated by randomly inserting fake objects of
known magnitude into each image and then recovering them with the same
detection parameters used for real objects.  The inserted objects were
given Gaussian profiles with a FWHM of 1\farcs3 to approximate the shape
of slightly extended, distant galaxies.  The resulting 80\% completeness
values in the \kband~are 22.5, 22.8, and 22.4 AB, which are similar to
the 5-$\sigma$ detection limit in each image.  

Infrared imaging is publicly available for GOODS-S.  We utilized
\kband~data taken with the SOFI instrument on the NTT because it is
similar in depth and resolution to the Palomar observations.  The SOFI
data reach $K_{s} < 22.5$ (AB, 5-$\sigma$) with stellar FWHM values less
than 1\farcs0 on average.  Detailed information on the SOFI observations
can be found in \citet{vandame01}.

For the infrared imaging in both GOODS fields, we used SExtractor
\citep{ber96} to make a K-band catalog, limited to $K_{s} = 22.4$ (AB),
and for total magnitudes we use the SExtractor Kron estimate.  We do not
adjust the Kron magnitude to account for missing light in extended
sources.  The optical-infrared color measurements, which are used for
estimating photometric redshifts and stellar masses, are determined from
1\farcs0 radius aperture photometry.  In the case of the infrared data,
these measurements are made by SExtractor.  They are then compared to
the catalog of optical ACS $BViz$ 1\farcs0 photometry as tabulated by
the GOODS team \citep{gia04}.

\subsection{ACS Morphologies}\label{morph}

Based on version 1.0 HST/ACS data released by the GOODS team
\citep{gia04}, a $z$-selected catalog was constructed with a magnitude
limit of $z_{AB} = 22.5$, where reliable visual morphological
classification was deemed possible.  The resulting sample of 2978
objects spread over both GOODS fields was inspected visually by one of us
(RSE) who classified each object \citep[using techniques discussed in
detail by][]{B98} according to the following scale: -2=Star, -1=Compact,
0=E, 1=E/S0, 2=S0, 3=Sab, 4=S, 5=Scd, 6=Irr, 7=Unclass, 8=Merger,
9=Fault.  Following \citet{BE00}, we divide these classes into three
broad categories: `E/S0' combines classes 0, 1, and 2 and contains 627
galaxies; `Spirals' combines classes 3, 4, and 5 and contains 1265
galaxies; and `Peculiar/Irregular' comprises classes 6, 7 and 8 and
contains 562 galaxies.  The morphological catalog is publicly available
at \verb+www.astro.caltech.edu/GOODS_morphs/+.

Some caution is required in comparing these morphological
classifications over a range of redshifts.  Surface brightness dimming
can bias high redshift morphologies towards early-type classifications.
We see evidence for this effect when we compare the 5 epoch, version-1.0
GOODS morphologies to previous determinations made by RSE for the same
sample of objects in a single epoch of the version-0.5 GOODS release.
The overall agreement is excellent, with the shallower single visit
classes offset from the deeper stacked equivalent by only 0.1 types on
the 12-class scheme defined above \citep[see][]{treu05}.  The standard
deviation of this comparison is 1.3 morphological classes, demonstrating
that the effect on the broader morphological categories, which combine
three classes into one, would be minimal.  Furthermore, the $z_{AB} <
22.5$ limit is one magnitude brighter than the 80\% completeness limit
of the ACS data for objects with half-light radii less than 1\farcs0
\citep{gia04}.  Thus the high signal-to-noise of the $z_{AB} < 22.5$
sample ensures robust classifications to $z \sim 1$.

Wavelength dependent morphological $k$-corrections are often more
important for comparisons across different redshifts.  The morphological
classifications made here were carried out first in the $z$-band, in
which the lowest redshift bin at $z \approx 0.3$ samples rest-frame
$R$-band while the highest redshift bin at $z\approx 1$ samples
rest-frame $B$-band.  However, after the first pass, $Viz$ color images
were inspected and about 5\% of the sample was corrected (by
never more than one class) based on the color information.  In this way,
the galaxies suffering most from the morphological $k$-correction were
accounted for.

We can gain a quantitative estimate for the remaining $k$-correction
effect by referencing the ``drift coefficients'' tabulated in Table 4 of
\citet{B98} and extrapolating them from the $I$-band to our $z$-band
classifications.  We would expect very small $k$-corrections until the
highest redshift bin ($z \approx 1$) which is equivalent in wavelength
to the $z=0.7$ interval in \citet{B98}.  At this point, \citet{B98}
estimate that $\approx$13\% of Spirals are misclassified as Peculiars,
while the Ellipticals and Spirals exchange $\approx$25\% of their
populations, leaving the relative numbers nearly the same.

The results of \citet{B98} were based on automated classifications
carried out using the Asymmetry--Concentration plane, however, and may
overestimate the magnitude of the $k$-correction.  Many other groups
have also investigated this effect \citep[e.g.][]{kuch00, kuch01,
  wind02, pap03} and suggest milder results.  The
$k$-corrections are found to be most severe when optical morphologies
are compared to the mid-UV.  None of the classifications in our sample
were based on restframe mid-UV morphology.  Furthermore, most studies
have found that early-type spirals, with their mixture of star formation
and evolved stellar populations, show the most drastic changes between
red and UV wavelengths, while other types vary less because they are
either completely dominated by young stars (late-type disks, Peculiars)
or have little to no star formation at all (Ellipticals).  Indeed,
\citet{conselice05} compare visual classifications similar to those
presented here in WFPC $I$ and NICMOS $H$-band for 54 galaxies in HDF-N with $z < 1$
and find that only 8 disagree.  Most of these were labeled as
early-types in $H$-band and as early-type disks in $I$-band.  Based on
these various studies, we expect that the wavelength dependent
$k$-corrections remaining in the sample are important only at the
highest redshifts.  Even then they are likely to be small because the
$z$-band is still redward of the rest-frame UV.  Although we choose not
to explicitly correct for the morphological k-correction, we estimate
that statistically over the redshift range 0.5-1, where the bulk of the
evolutionary trends are seen, at most 5---10\% of the populations are
misidentified in our broad classification system.

\subsection{Spectroscopic and Photometric Redshifts}\label{photoz}

Accurate redshifts are important not only for identifying members of a
given redshift interval but for determining luminosities and stellar
masses.  Spectroscopic redshifts were taken from two sources.  The Keck
Team Redshift Survey (KTRS) provides redshifts for GOODS-N and was
selected in $R$ and carried out with DEIMOS on Keck II \citep{wirth04}.
The KTRS is 53\% complete to $R_{AB}<24.4$, giving a sample of 1440
galaxy redshifts and providing redshifts for 761 galaxies (58\%) in the
GOODS-N morphological catalog described above.  After our additional
\kband~requirement (see $\S$\ref{completeness}), this number reduces to
661 galaxies primarily because the \kband~imaging covers only
$\approx$70\% of GOODS-N.  Spectroscopic redshifts in GOODS-S were taken
from the VIMOS VLT Deep Survey \citep{lef04} which is 88\% complete to
$I_{AB} =24$ and accounts for 300 galaxies (25\%) in the GOODS-S
morphological catalog.  We supplement this sample with 792 photometric
redshifts (66\%) from COMBO-17 \citep{wolf04}.
 
\begin{figure}
\caption{Results of the photometric redshift estimation using the BPZ
  code by \citet{ben00}.  The plot illustrates the difference
  between photo-z's and spectroscopic redshifts (where they exist).\label{photz}}
\includegraphics[scale=1]{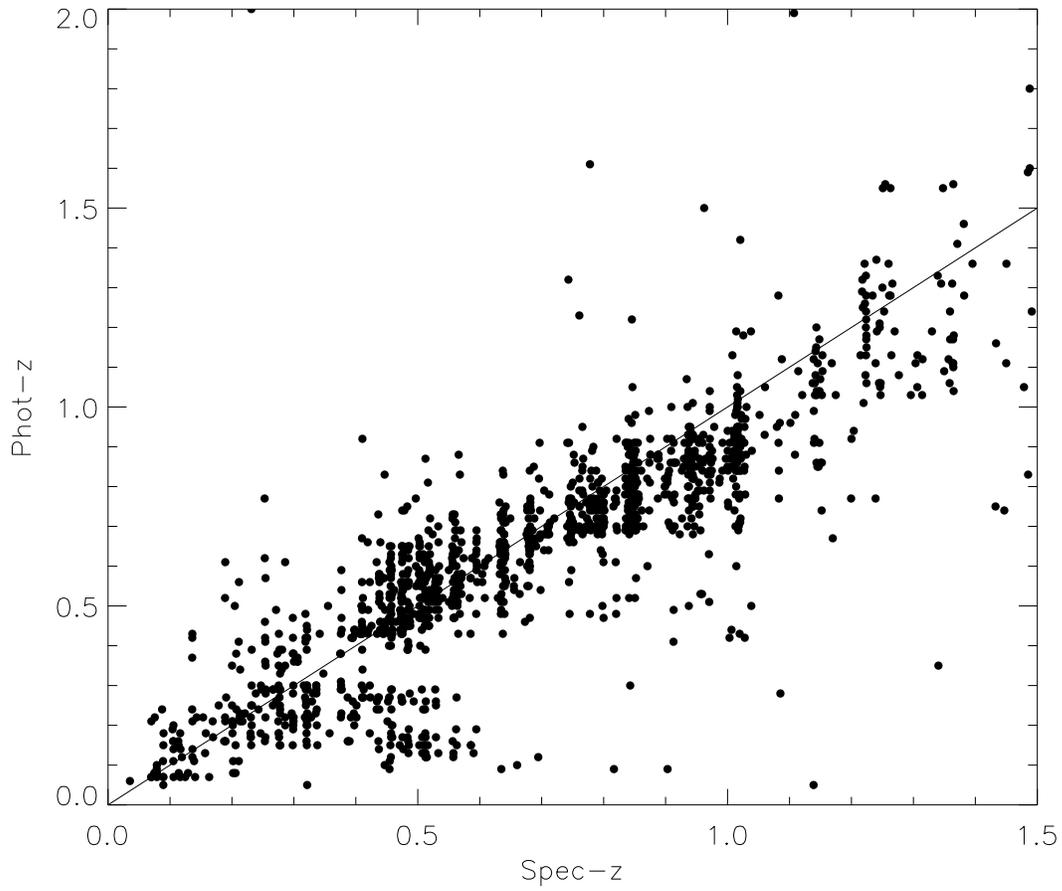}
\end{figure}

For more than half of the total morphological sample, published
redshifts are not available and photometric redshifts have to be
measured.  In GOODS-S, 107 photometric redshifts were needed while in
GOODS-N the number was 343.  The inclusion of these redshifts and the
requirement of \kband~photometry yields a final morphological sample
that is complete to $z_{AB} < 22.5$ and $K_{s} < 22.4$ (AB).  We also
constructed a fainter sample with $z_{AB} < 23.5$ that is too faint for
reliable morphological classification but allows for investigations of
completeness (see $\S$\ref{tot_mfn}).  Photometric redshifts were
estimated using the Bayesian Photometric Redshift (BPZ) Code described
in \citet{ben00}.  Using the same priors that \citet{ben00} applies to
the HDF-N, we ran the BPZ software on the 1\farcs0 diameter
fixed-aperture ACS and \kband~photometry, allowing for two interpolation
points between templates.  The KTRS and VIMOS spectroscopic redshifts
were used to characterize the quality of a subset of the photometric
redshifts (see Figure \ref{photz}).  The results were similar for both
GOODS fields, with a combined mean offset of $\Delta z
/ (1+z_{spec}) = -0.02$ and an rms scatter of $\sigma [\Delta z /
(1+z_{spec})] = 0.12$, similar to the precision achieved by \citet{Mo04}
who used BPZ to estimate redshifts in GOODS-S.  We did not, however,
correct for the poorer resolution ($\sim$1\farcs0) of the \kband~data.

The comparison between photo-$z$'s and spec-z's in Figure \ref{photz}
shows a tendency for photo-$z$'s to be underestimated.  There is a set
of objects with spec-$z < 0.5$ that have photometric redshifts near 0.2.
At spec-$z \approx 1$ there is another more mild deviation toward lower
photo-$z$'s that is likely due to the Bayesian prior \citep{ben00} which
assumes a decreasing redshift distribution at $z \sim 1$.  And in
general, there appears to be more catastrophic outliers with photo-$z$
underestimates.  Much of this behavior is likely related to the lack of
U-band photometry, which is crucial to ruling out false low-$z$
photometric solutions.

Over the redshift range of interest, $0.2 < z < 1.4$, the final sample
consists of 1906 galaxies, of which 893 (47\%) have spectroscopic
redshifts and 1013 (53\%) have photometric redshifts.  We divide the
sample into three redshift intervals, $0.2 < z < 0.51$, $0.51 < z <
0.8$, and $0.8 < z < 1.4$, chosen to balance the number of objects in
each interval.  The morphological breakdown (Ellipticals, Spirals,
Peculiars) in each redshift bin is as follows: for $0.2 < z < 0.51$,
(167, 353, 126); for $0.51 < z < 0.8$, (220, 326, 138); and for $0.8 < z
< 1.4$ (131, 291, 154).

\section{Determination of Stellar Masses}\label{mass_code}

\begin{figure}
\caption{Examples of the best-fitting model spectra and the resulting
  stellar mass probability distribution as determined by the stellar
  mass code.  The photometry points are plotted and final mass
  indicated.  The dashed lines denote the 68\% confidence intervals in
  the derived stellar mass. \label{demofit}} \plotone{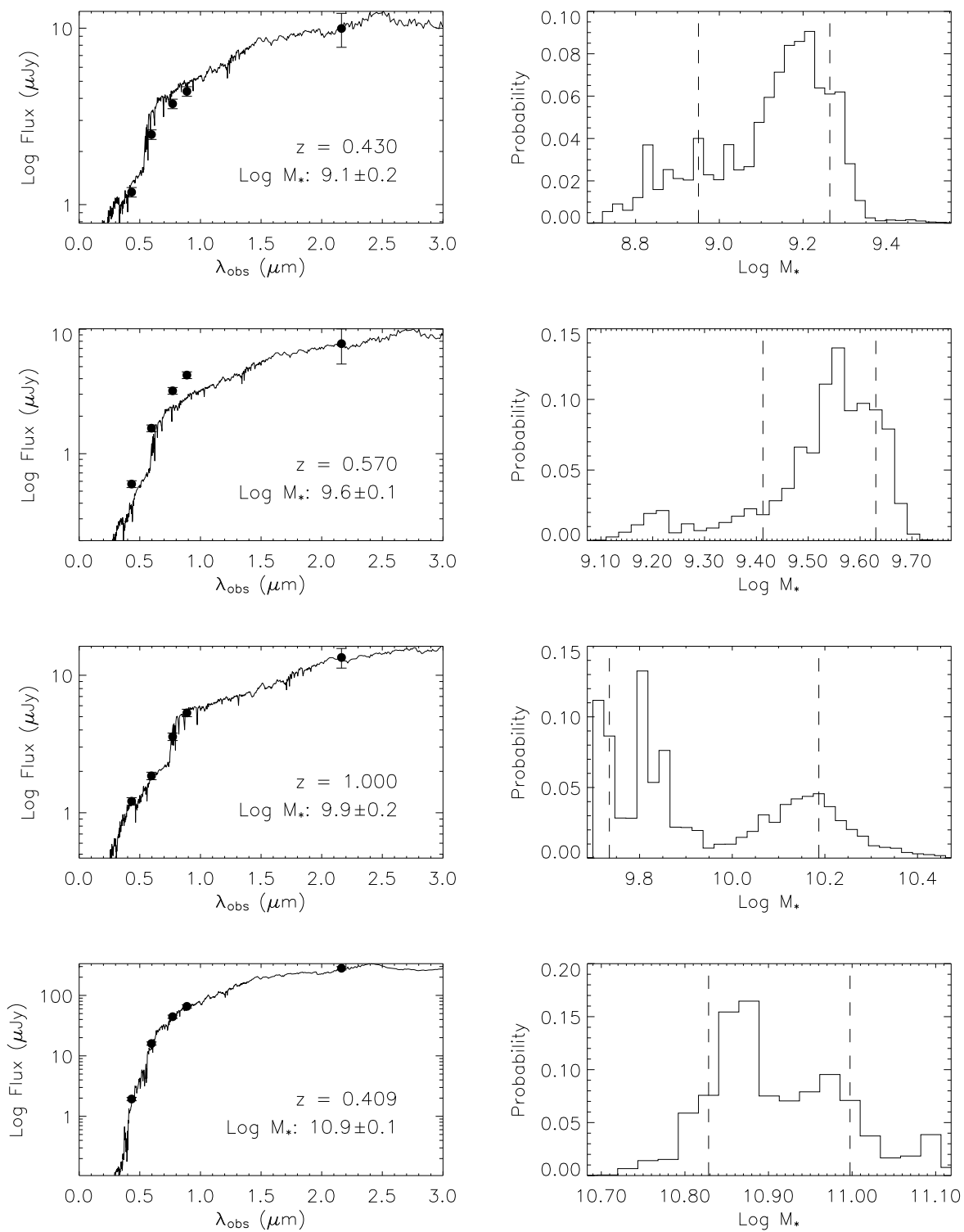}
\end{figure}

Estimating stellar masses using the combination of infrared imaging,
multiband photometry, and redshift information is now a widely applied
technique first utilized by \citet{BE00}.  In this paper, we use a new
Bayesian stellar mass code based on the precepts described in
\citet{kauff03a}.  Briefly, the code uses the multiband photometry and
redshift to compare the observed SED of a sample galaxy to a grid of
synthetic SEDs \citep[from][]{BC03} spanning a range of star formation
histories (parameterized as an exponential), ages, metallicities, and
dust content.  The $K$-band mass-to-light ratio ($M_*/L_K$), stellar
mass, minimum $\chi^2$, and the probability the model represents the
data is calculated at each grid point.  The probabilities are then
summed across the grid and binned by stellar mass, yielding a stellar
mass probability distribution for each galaxy (see Figure
\ref{demofit}).  We use the median of the distribution as an estimate of
the final stellar mass.  Photometry errors enter the analysis by
determining how well the model SEDs can be constrained by the data.
This is reflected in the stellar mass probability distribution which
provides a measure of the uncertainty in the stellar mass estimate given
by the width of the distribution.

In addition to accounting for photometric errors, the mass probability
distribution also includes the effects of mass degeneracies in the model
parameter space.  These two effects typically account for 0.2 dex of
uncertainty.  In addition, systematic errors enter through assumptions
about model parameters.  In some cases, the observed SEDs do not fall in
the parameter space spanned by the grid.  These are characterized by
large minimum $\chi^2$ values and we add an additional 0.2 dex in
quadrature to their uncertainties.  The largest systematic source of
uncertainty comes from our assumed IMF, in this case that proposed by
\citet{chabrier03}.  Masses derived assuming this IMF can be converted
to Salpeter by adding 0.3 dex.  Various tests of the stellar mass code
used here under different assumptions for the IMF, the model parameter
space, and the star formation parameterization are presented in
\citet{bundy05}.

Despite the large number of spectroscopic redshifts, when the
stellar mass estimator is applied to galaxies with photometric
redshifts, additional errors must be included to account for the much
larger redshift uncertainty.  At the same time, catastrophic photo-z
errors (which are apparent in Figure \ref{photz}) can significantly
affect mass estimates.  Redshift errors enter in two ways.  First, they
affect the determination of the galaxy's restframe SED because
$k$-corrections cannot be as accurately determined.  This can alter the
best fitting model and the resulting mass-to-light ratio.  Far more
important, however, is the potential error in the luminosity distance
from the increased redshift uncertainty.  For the standard cosmology we
have assumed, a redshift uncertainty of $\sigma [\Delta z /(1+z)] =
0.12$ can lead to an error of roughly 20\% in luminosity distance.  This
can contribute to an added mass uncertainty of almost 50\%.

\begin{figure}
\caption{Difference in estimated stellar mass for our spectroscopic
  sample when photo-z's are used instead of spec-z's.  The shaded region
  shows the expected standard deviation resulting from variations in the
  luminosity distance due to photo-z error. \label{phot_mass}}
  \plotone{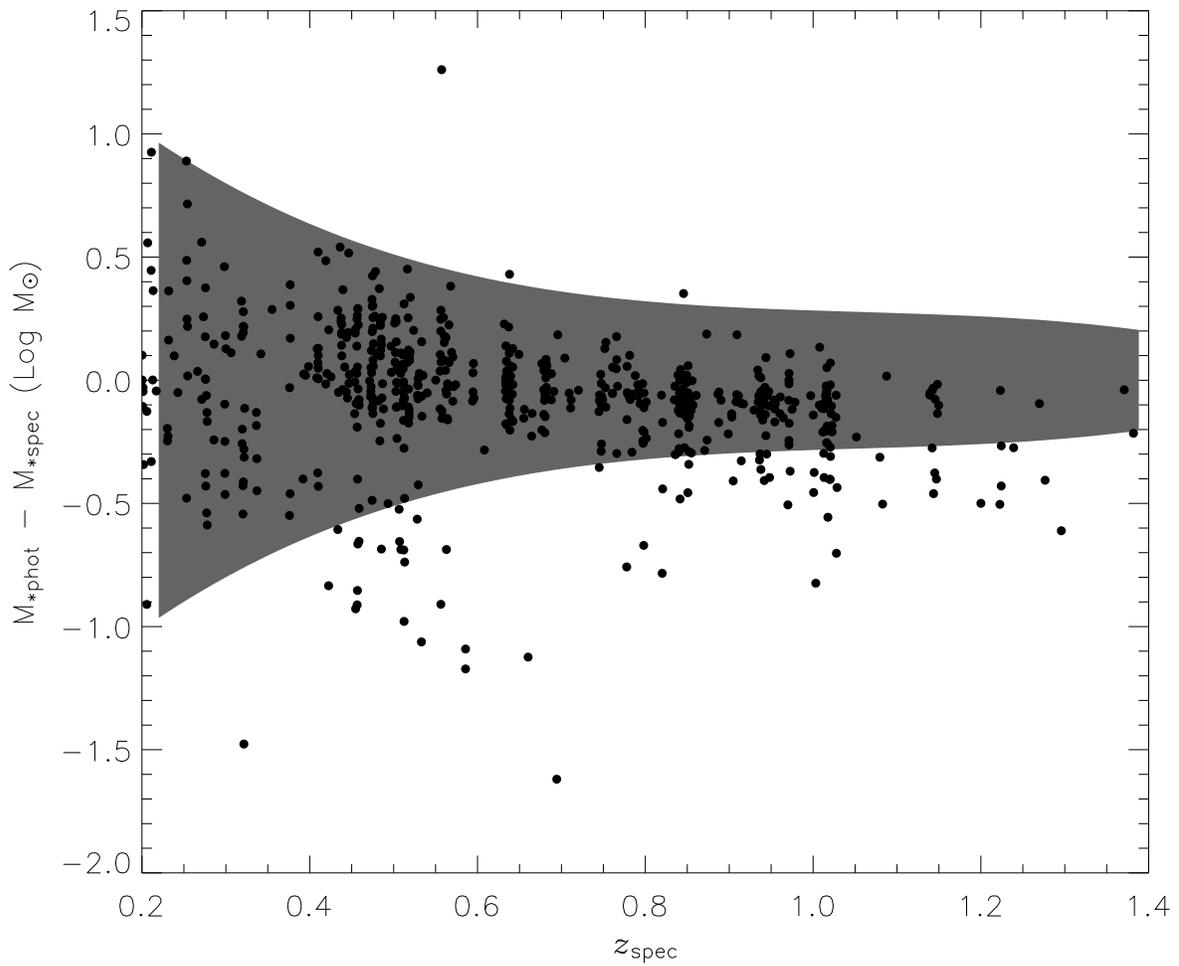}
\end{figure}

The additional stellar mass uncertainty resulting from the use of
photo-z's is illustrated in Figure \ref{phot_mass}.  In this experiment,
we measure photometric redshifts for galaxies that already have secure
spectroscopic redshifts and use these photometric redshifts to determine
a second set of stellar mass estimates.  Figure \ref{phot_mass} shows
the difference in stellar mass for the same galaxies when photometric
redshifts are used instead of spectroscopic redshifts, plotted as a
function of their spectroscopic redshift.  Individual mass
estimates become less certain and there are several catastrophic
outliers with stellar masses that differ by an order of magnitude.  The
shaded region shows the standard deviation in stellar mass error based
on a Monte Carlo simulation of 20,000 galaxies in which simulated
redshifts were drawn from the observed photometric redshift error
distribution ($\sigma [\Delta z /(1+z)] = 0.12$).  The simulation includes only
the primary effect on the luminosity distance.  The shaded region
accounts for both the rms uncertainty and the effect of catastrophic
photo-z failures since both are included in the measurement of $\sigma
[\Delta z /(1+z)]$.  

Figure \ref{phot_mass} shows a systematic offset such that most of the
dramatic outliers tend to have lower masses when photometric redshifts
are used than when spectroscopic redshifts are used.  This trend can be
understood by comparing to Figure \ref{photz}, which, as discussed in
$\S$\ref{photoz}, shows a majority of outliers with lower photo-z
measurements as compared to their spectroscopic values.  The smaller
luminosity distance that results from the photo-z underestimate leads to
stellar masses that are also underestimated.  Although we recognize this
systematic trend, we do not attempt to correct for it since it primarily
affects the outliers.  We do add an additional 0.3 dex of uncertainty to
stellar masses gleaned from the photometric redshift sample.

\section{Completeness and Selection Effects in the Sample}\label{completeness}

\begin{figure}
\caption{$(z-K_s)$ versus $K_{s}$ color-magnitude relation.  The
  small solid points are galaxies with spectroscopic redshifts.  Open
  symbols denote those galaxies with photometric redshifts.  The thin solid line illustrates
  the $z_{AB} < 22.5$ morphological classification limit, while the dotted line traces the
  completeness limit of the \kband~data from Palomar and ESO.  Three simple stellar
  population models, each with $L_K^*$ luminosity and a
  formation redshift of $z_{form} = 10$, are also plotted.  Models A and
  B have exponential SF timescales of $\tau = 0.4$ Gyr, and
  metallicities of Z=0.05 for A and Z=0.02 (Z$\odot$) for B.  Model C
  has $\tau = 4.0$ Gyr and solar metallicity.  The large solid dots
  denote redshifts, from left to right, $z=$ 0.4, 0.8, and 1.2.
  \label{zkcut}}
\plotone{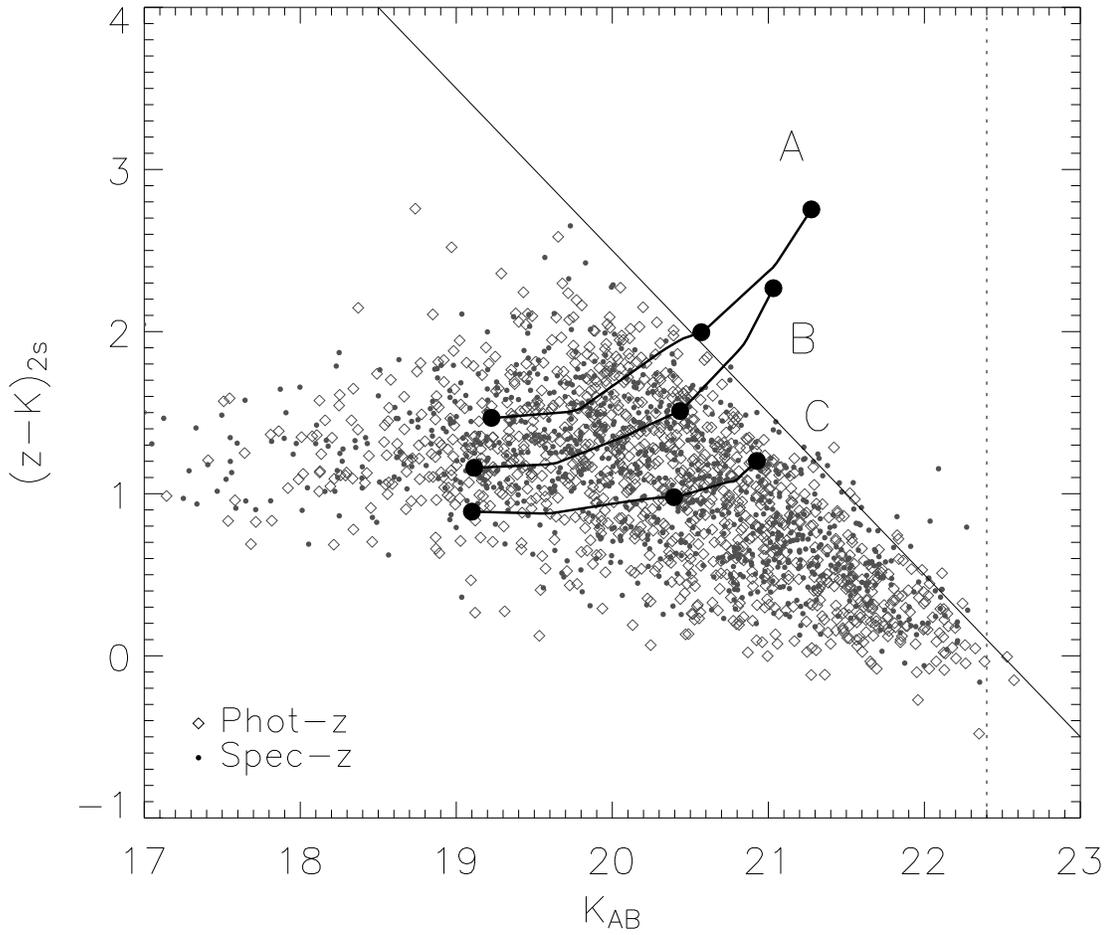}
\end{figure}

The final sample combines several datasets leading to complicated
completeness and bias effects that must be carefully examined.  First,
as described in $\S$\ref{morph}, because accurate morphological
classifications are required, the sample was limited to $z_{AB} < 22.5$
to ensure the fidelity of those classifications.  Second, reliable
stellar mass estimates at redshifts near $z \approx 1$ require three key
ingredients: 1) Multi-band optical photometry, 2) \kband~photometry, and
3) Redshifts.  The optical photometry comes from the GOODS ACS imaging.
Since the ACS catalog was selected in the $z$-band, the $z_{AB} < 22.5$
limit applies to these data as well.  The \kband~imaging was described
in $\S$\ref{kimaging}.  As illustrated in the color-magnitude diagram in
Figure \ref{zkcut}, this depth is adequate for detecting the vast
majority (95\%) of the objects that lie within the area covered by the
infrared imaging and satisfy the $z_{AB} < 22.5$ criterion.  Thus, with
the sample already limited in the $z$-band, requiring an additional
\kband~detection does not introduce a significant restriction and we can
consider the final sample complete to $z_{AB} < 22.5$.  

The third ingredient in the stellar mass estimate---the galaxy's
redshift---comes from a combination of sources ($\S$\ref{photoz}).
Figure \ref{zkcut} plots the location on the color-magnitude diagram of
those galaxies with $z_{AB} < 22.5$ that do not have spectroscopic redshifts
(open symbols).  They account for 54\% of the final sample and are
assigned redshifts based on the photo-z technique described in
$\S$\ref{photoz}.  The fact that almost half of the galaxies in the sample
have spectroscopic redshifts is an important advantage for precise
mass functions.  Relying entirely on photo-z's would  blur the
edges of the redshift intervals and introduce additional uncertainty
to every galaxy in the survey, as discussed in $\S$\ref{mass_code}.


Despite the limitations of the various datasets required for this study,
the final sample suffers almost exclusively from the magnitude cut in
the $z$-band.  Because it is a magnitude limited sample, it is
incomplete in mass and at the highest redshifts, the objects with the
reddest $(z - K_s)$ colors will begin to drop out of the sample.  This
is illustrated by the expected location on the color-magnitude diagram
of three different stellar population models, each with a luminosity of
$L_K^* \approx -24$ at all redshifts (see Figure \ref{zkcut}).  Models A
and B have exponential SF timescales of $\tau = 0.4$ Gyr, and
metallicities of Z=0.05 ($2.5 Z\odot$) for A and Z=0.02 (Z$\odot$) for
B.  Model C has $\tau = 4.0$ Gyr and solar metallicity.  The large solid
dots denote redshifts $z =$ 0.4, 0.8, and 1.2.  In the highest redshift
bin ($0.8 < z < 1.4$), the redder, passively evolving sources can be
expected to suffer most from the $z$-band cut.  As these are likely to
be massive, we would expect their absence to also be reflected in the
combined mass function.

Following previous work \citep[e.g.][]{fontana03}, we can translate the
$z$-band cut into a conservative mass completeness limit by estimating
the mass corresponding to a reasonable maximum $M_*/L_z$ ratio as a
function of redshift.  To do this, we calculate the mass of a
near-instantaneous burst model with a formation redshift of $z_{form} =
10$, no dust, sub-solar metallicity, and a luminosity corresponding to
an observed magnitude of $z_{AB} = 22.5$ at all redshifts.  The
resulting mass completeness limits rise from $10^{10}$ \msun~at $z \sim
0.3$ to $10^{11}$ \msun~at $z\sim 1$.  In $\S$\ref{morph_fn}, we further
discuss how mass incompleteness impacts the derived mass functions at
$0.8 < z < 1.4$.

\section{Results}\label{results}

\subsection{Methods and Uncertainties}\label{methods}

Given the small area of the GOODS fields (0.1 square degrees), cosmic
variance and clustering in these intervals will affect the mass
functions we derive.  \citet{som04b} present a convenient way to
estimate cosmic variance based on the number density of a given
population and the volume sampled.  Using these techniques, we estimate
that uncertainties from cosmic variance range from $\approx$20\% in the
highest redshift bin to $\approx$60\% in the lowest.  This translates
into an additional 0.1--0.3 dex of uncertainty in the final mass
functions.

In deriving a mass or luminosity function in a magnitude limited survey,
faint galaxies not detected throughout the entire survey volume must be
accounted for.  Many techniques exist to accomplish this while
preventing density inhomogeneities from biasing the shape of the derived
luminosity function \citep[for a review, see][]{willmer97}.  However,
there is no cure for variations from clustering and cosmic variance.  With an expected uncertainty from
cosmic variance of $\sim$40\% on average, these variations will affect
comparisons we might draw between redshift intervals.  Given these
limitations we adopt the simpler $V_{max}$ formalism \citep{schmidt68}.
The $V_{max}$ estimate is the volume corresponding to the highest
redshift, $z_{max}$, at which a given galaxy would still appear brighter
than the $z_{AB} = 22.5$ magnitude limit and would remain in the sample.
For a galaxy $i$ in a redshift interval, $z_{low} < z < z_{hi}$,

\begin{equation}
V^i_{max} = d \Omega \int_{z_{low}}^{{\rm min}(z_{hi}, z_{max})} \frac {dV}{dz} dz
\end{equation}

\noindent where $d \Omega$ is the solid angle subtended by the survey
and $dV/dz$ is the comoving volume element.  

The stellar mass estimator fits a model spectrum to each galaxy.  By
redshifting this model spectrum and integrating it over the $z$-band filter response
function, we can calculate the apparent $z$-band magnitude as a function
of redshift, implicitly accounting for the $k$-correction.  The
quantity, $z_{max}$, is determined by the redshift at which the apparent
magnitude becomes fainter than $z_{AB} = 22.5$.

Once $V_{max}$ is estimated, we calculate the comoving number density of
galaxies in a particular redshift bin and stellar mass interval, $(M_* +
dM_*)$, as,

\begin{equation}
\Phi(M_*) dM_* = \sum_i \frac {1}{V^i_{max}} dM_*
\end{equation}

\noindent where the sum is taken over all galaxies $i$ in the interval.

\begin{figure}
\caption{Redshift distributions for the primary GOODS sample with $z_{AB} < 22.5$. \label{zdist}}
\plotone{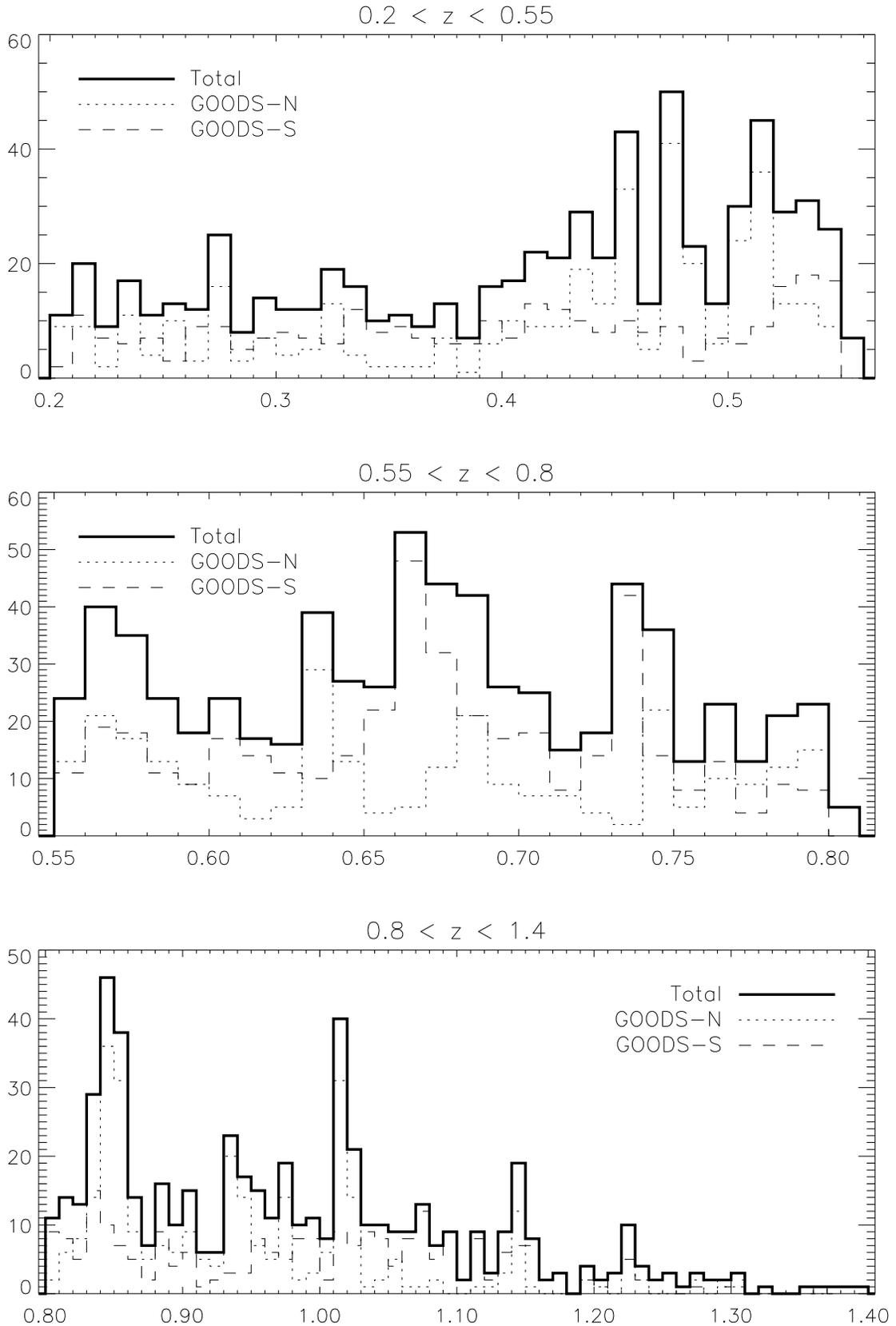}
\end{figure}

The $V_{max}$ formalism is appealing because it is easy to apply and
makes no assumptions on the form of the luminosity or mass function.  It
can be biased by clustering since it assumes galaxies are uniformly
distributed through the survey volume.  This bias is best understood
through the observed redshift distributions (Figure \ref{zdist}).  In
the lowest redshift bin, the concentration at the high end of the
redshift interval near $z=0.47$ leads to an underestimate in the mass
function, especially for the number densities of fainter galaxies that
cannot be detected at the furthest distances in this interval.  In the
highest redshift bin, the redshift spikes at the low-$z$ end have the
opposite effect.  In this case, the clustering increases the resulting
mass function by causing an overestimate of galaxies that would not be
detected if the redshift distribution was more uniform.  Clustering can
also affect the type-dependent stellar mass functions because early-type
galaxies are expected to be more strongly clustered than later types.

Other uncertainties in deriving the mass function are greatly reduced by
utilizing spectroscopic redshifts.  The effect on the mass functions
caused by uncertainty in photo-z estimates and stellar mass errors is
estimated in the following way.  We use a Monte Carlo technique to
simulate 100 realizations of our dataset, utilizing the resulting
variation in the observed mass functions to interpret the errors.  For a
given realization, the stellar mass of each galaxy is drawn randomly
from the stellar mass probability distribution determined by the mass
estimator (see $\S$\ref{mass_code}); thus avoiding any assumption about
the form of the estimated mass distribution.  Galaxies with photometric
redshifts tend to smear out the edges of the redshift intervals.  In the
simulations, the realized redshift for these galaxies is drawn from a
Gaussian distribution with $\sigma = \sigma [\Delta z / (1+z_{spec})] =
0.12$, the same rms measured for galaxies with both spectroscopic and
photometric redshifts.  The effect of this redshift uncertainty on the
luminosity distance is also included in the stellar mass error budget
(see $\S$\ref{mass_code}).

\subsection{Galaxy Stellar Mass Functions}\label{tot_mfn}

\begin{figure}
\caption{Total galaxy stellar mass functions in three redshift
intervals.  The solid line is the best fit to the $z=0$ mass function
from \citet{cole01}.  Results from \citet{drory04} and the ``best--fit''
stellar mass functions from \citet{fontana04} are shown, after
correcting for different choices of the IMF and Hubble Constant.  For
the brighter, $z_{AB} < 22.5$ sample, mass completeness limits based on
a maximum reasonable $M_*/L_K$ ratio (see $\S$\ref{completeness}) are
indicated by the dotted lines.  The error bars are calculated from Monte
Carlo simulations that account for uncertainties arising from the
photometry, stellar mass estimates, photo-$z$ estimates, and Poisson
statistics.
\label{totmassfn}} \includegraphics[scale=0.9]{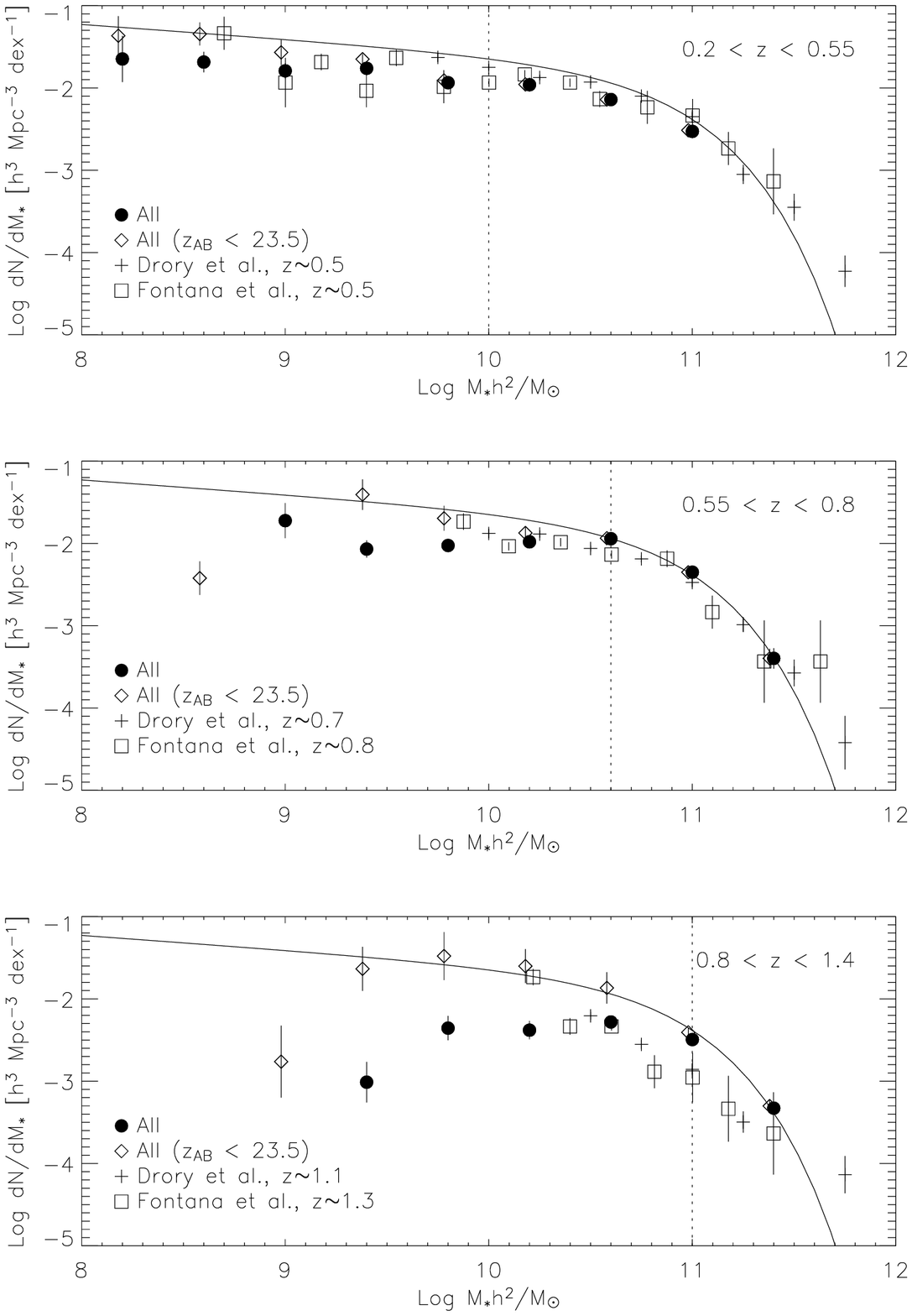}
\end{figure}

We plot the resulting galaxy stellar mass functions for all types in
Figure \ref{totmassfn}.  The solid lines trace the best fit to the local
mass function measured by \citet{cole01} (we do not plot the results of
\citet{bell03} which are consistent with \citet{cole01}).  The redshift
dependent mass functions derived by \citet{drory04} (also plotted as
``plus'' symbols) come from a larger photo-z sample that is most
complete at higher masses.  Results from the spectroscopic K20 survey
are shown as squares \citep{fontana04}.  We note that the lower result
from \citet{fontana04} in the high-$z$ bin may be caused by mismatched
redshift intervals.  Their result for $1.0 < z < 1.5$ is plotted, but
their $0.7 < z < 1.0$ mass function (shown in the middle plot of Figure
\ref{totmassfn}) may provide a more adequate comparison considering our
high-$z$ bin includes many galaxies in the range, $0.8 < z < 1.0$.

Also plotted in Figure \ref{totmassfn} are the combined mass functions
for a larger, unclassified, photometric sample where the $z$-band
magnitude cut has been increased to $z_{AB} = 23.5$.  Though reliable
visual morphological classification is not feasible for objects in the
ACS data with $z_{AB} > 22.5$, the fainter sample demonstrates the
effects of incompleteness in the primary ($z_{AB} = 22.5$) catalog,
which become particularly important in the highest redshift bin.  The
point at which the morphological sample begins to show a deficit with
respect to the fainter sample is consistent with the mass completeness
limits calculated based on the maximum $M_*/L_z$ ratio (see
$\S$\ref{completeness}).  For the three redshift intervals, the
estimated mass incompleteness limits are, in order of increasing
redshift, $10^{10}$ \msun, $4 \times 10^{10}$ \msun, and $10^{11}$
\msun.  As discussed in $\S$\ref{methods}, in addition to completeness,
cosmic variance and clustering dominate the small GOODS area and make
interpretations of the mass function difficult.
 
We fit Schechter functions to the binned data, including fits to
separate morphological populations (see $\S$\ref{morph_fn}), and show
the resulting parameters in Table \ref{schfit}.  The primary and
extended samples are quite similar in the first two redshift intervals,
with incompleteness in the primary sample becoming significant in the
third ($0.8 < z < 1.4$).  Across the full redshift range, the combined
mass function shows little evolution and remains similar in shape to its
form at $z=0$.  

These results are consistent with previous work.  \citet{fontana04} find
little evolution out to $z \sim 1$ in the mass function derived from the
K20 survey (also plotted).  The K20 sample \citep{cimatti02} has good
spectroscopic coverage (92\%) but is slightly shallower (0.5 mag in $K$)
and roughly one quarter of the size of the sample studied here.  The
MUNICS survey \citep{drory01}, in contrast, contains 5000 galaxies
spread over several fields, covering a much larger area.  It is roughly
one and a half magnitudes shallower in $K$ and consists primarily
($\approx$90\%) of photometric redshifts.  \citet{drory04} argue that
the MUNICS combined mass function exhibits significant evolution to
$z=1.2$, with a decrease in the characteristic mass and steepening of
the faint--end slope.  Considering the uncertainties, we feel the MUNICS
result is consistent with the work presented here although we differ in
the interpretation.

Theoretical models currently yield a variety of predictions for the
combined galaxy stellar mass functions.  \citet{fontana04} compare
predictions from several groups with various techniques including
semi--analytic modeling and numerical simulations.  In general,
semi--analytic models \citep[e.g.][]{cole00, som01, menci04} tend to
produce mass functions that evolve strongly with redshift, with
decreasing normalization and characteristic masses that together
underpredict the observed number of massive, evolved galaxies at high
redshift.  Rapid progress in addressing this problem is currently
underway.  Numerical models \citep[e.g.][]{nag04} better reproduce a
mildly evolving mass function and more easily account for high mass
galaxies at $z\sim 1$, but sometimes overpredict their abundance (see
the discussion in $\S$\ref{morph_fn} on massive galaxies).  Observations
of the galaxy stellar mass function and its evolution thus provide key
constraints on these models.

\subsection{Type-Dependent Galaxy Mass Functions}\label{morph_fn}

In Figure \ref{morph_mfn}, we show the galaxy stellar mass functions for
the three broad morphological populations derived using the $V_{max}$
formalism described in $\S$\ref{methods}.  As a check, we also
calculated mass functions after applying a $z$-band absolute magnitude
limit of $M_z < -20.3$, to which nearly every galaxy can be detected at
every redshift.  Absolute magnitudes are determined through SED fitting
as part of the stellar mass estimator ($\S$\ref{methods}). Though less
complete, the general characteristics of the mass functions with $M_z <
-20.3$ agree well at high stellar masses with those derived using the
$V_{max}$ method.

All three redshift bins in Figure \ref{morph_mfn} are complete at $M_* >
10^{11}$ \msun, and little difference is seen in the combined mass
function between the morphological sample (with $z_{AB} < 22.5$) and the
fainter sample ($z_{AB} < 23.5$).  This allows for a comparison of the
morphological makeup of the high-mass population.  In the highest
redshift bin, we find that Spirals are slightly favored at $M_* \approx
10^{11}$ \msun~and are competitive with E/S0's at $M_* \approx 3 \times
10^{11}$ \msun.  At the same time, Peculiars make a significant
contribution at these masses.  Both the Peculiar and Spiral populations
drop at lower redshifts as E/S0's become increasingly dominant at high
masses.

Turning now to the broader range of masses sampled by the data, the
first redshift bin exhibits a transitional mass of $M_{tr} \approx
2$--$3 \times 10^{10}$ \msun~($\log_{10} M_{tr} \approx 10.3$--$10.5$)
below which the E/S0 population declines while the Spiral population
rises, becoming the dominant contributor to the combined mass function.
Across the three redshift bins studied here, it appears this
transitional mass shifts to lower mass with time as the contribution of
the E/S0 population to low mass galaxies increases.  We caution that
$M_{tr}$ is close to the estimated mass completeness limit, especially
in the $0.55 < z < 0.8$ redshift bin.

Tracing evolution at masses below $M_{tr}$ is difficult because of
incompleteness.  As discussed in $\S$\ref{completeness}, we would expect
early-type galaxies to be increasingly absent in the $z$-selected sample
at higher redshifts.  Thus, much of the more rapid decline in the E/S0
contribution at $0.8 < z < 1.4$ may be due to the $z$-band magnitude
cut.  When our cut is relaxed to $z_{AB} = 23.5$, we can expect that
many of these missing E/S0's will re-enter the sample and that this
would drive the combined mass function to levels more comparable to
those observed at lower redshift.  One line of evidence in support of
this is the predominantly red $(V - K_s)$ color of the galaxies
introduced into the sample when the $z_{AB}$ limit is relaxed to 23.5.
For stellar masses greater than $\approx$10$^{10}$ \msun, the $(V -
K_s)$ distribution of these galaxies is very similar to the E/S0
population in the $z_{AB} < 22.5$ sample.  Their low asymmetries, as
measured through the CAS system \citep{conselice03}, are also consistent
with an E/S0 population.  With decreasing stellar mass, the asymmetry
values increase and the color distribution spreads towards the blue,
suggesting that, at lower masses, other galaxy types enter the fainter,
high-$z$ sample as well.

Our suggestion that an E/S0 population is primarily responsible for
adjusting the combined mass function (for $M_* \gtrsim 10^{10}$ \msun)
is also consistent with the contribution of Spirals which, in the
high-$z$ bin, is a factor of $\sim$70\% lower at $M_* = 1.6 \times
10^{10}$ \msun~ ($\log_{10} M_*/M_{\odot} = 10.2$) than in the mid-$z$
bin.  At this same mass in the high-$z$ bin, however, the combined mass
function increased by almost an order of magnitude when galaxies with
$z_{AB} < 23.5$ are included.  Much of this increase is likely to come
from E/S0's that were previously missed.

\subsection{Integrated Stellar Mass Density}\label{int}

Following \citet{BE00}, in Figure \ref{rhoz} we show results for the
type-dependent stellar mass density as a function of redshift.  In this
analysis we have implemented a mass limit of $M_* > 10^{11}$
\msun, thus we are mass complete in all three redshift bins.  We find
that the total stellar mass density grows by $\simeq$30\% from $z \sim
1$ to $z=0$, although the uncertainty from cosmic variance is large.
The morphological breakdown of the increase in stellar mass density
illustrates a demise in the contribution from Spirals and Peculiars
accompanied by a rise in the density of Ellipticals.  The error bars
indicated on Figure \ref{rhoz} were calculated using the same Monte
Carlo technique that was applied to the mass functions.  In this way,
the uncertainties account for errors in photometry, stellar mass
estimates, and photometric redshifts.  The shaded region in Figure
\ref{rhoz} demonstrates the expected uncertainty due to cosmic variance.

The observed 30\% increase in the integrated stellar mass density is
consistent with measurements by other groups \citep[e.g.][]{cohen02,
fontana03, dickinson03} and can be reconciled with the observed star
formation rate over the same redshift range \citep[e.g.][]{fontana03}.
As shown in Figure \ref{rhoz}, we find similar results for the
morphological dependence of the growth of stellar mass as \citet{BE00}.
The sample presented here is more complete because it was selected in
$z$-band not $I$-band and is almost 7 times larger, resulting in smaller
random errors.  We note that \citet{BE00} do not estimate the affect of
cosmic variance on their sample, which we find to be the primary
uncertainty, nor do they ensure their sample is mass complete.  As
suggested by \citet{BE00}, the rise of the Elliptical population implies
that other galaxies, including Peculiars and Spirals, transform into
early-type galaxies with time.

\section{Discussion}

In broad terms, our results support two independent and consistent
perspectives on the mass assembly history of galaxies since $z \sim 1$.
On the one hand, global measures of the mass distribution, such as the
combined galaxy stellar mass function, show little evolution in the
distribution of massive galaxies from $z \sim 1$ to $z=0$.  This implies
that much of the observed star formation and associated stellar growth
over this interval occurs in lower mass galaxies.  On the other hand, a
more dynamic perspective emerges when the mass distribution is
considered according to galaxy type.  In this paper, we have shown that
although the number density of massive galaxies is relatively fixed
after $z \sim 1$, the morphological composition of this population is
still changing, such that the more balanced mix of morphological types
at $z \sim 1$ becomes dominated by ellipticals at the lowest redshifts.
Strikingly, this morphological evolution also appears to occur first at
the highest masses, proceeding to rearrange the morphologies of lower
mass galaxies as time goes on.  In this section, we explore how these
two perspectives---little total stellar mass evolution overall
accompanied by more substantial internal changes---lend insight into the
mechanisms responsible for the growth of galaxies and the development of
the Hubble Sequence.

\begin{figure}
\caption{Galaxy stellar mass function in three redshift intervals split
  by morphology.  The $z=0$ mass function from \citet[][solid
  line]{cole01} and the total mass functions are also shown.  Mass
  completeness limits based on a maximum reasonable $M_*/L_K$ ratio (see
  $\S$\ref{completeness}) are indicated by the dotted lines.  The error
  bars are calculated from Monte Carlo simulations that account for
  uncertainties arising from the photometry, stellar mass estimates,
  photo-$z$ estimates, and Poisson statistics.\label{morph_mfn}}
  \includegraphics[scale=0.95]{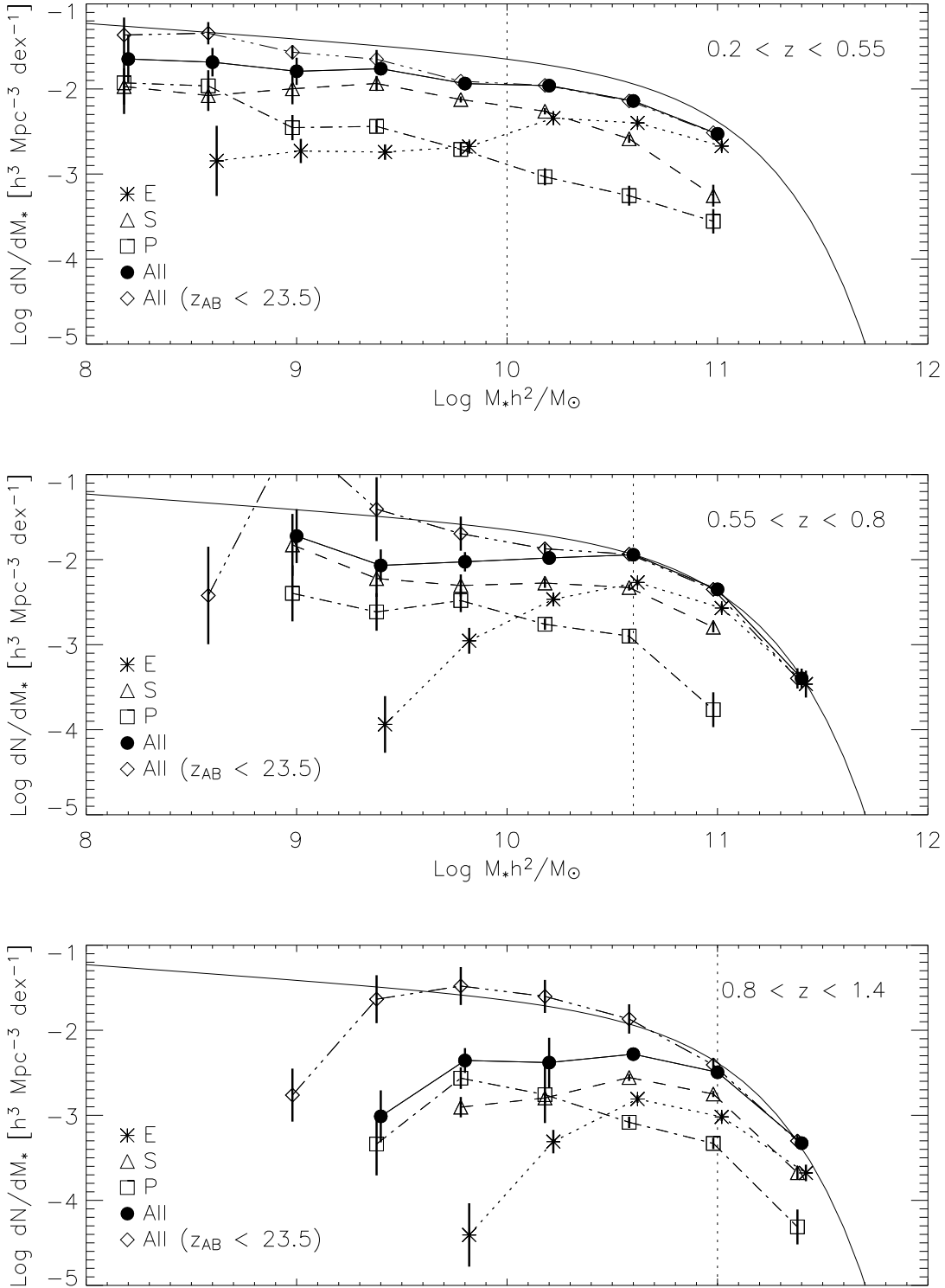}
\end{figure}

\begin{figure}
\caption{Integrated stellar mass density as a function of redshift,
  split by morphology and with a mass cut of $M_* > 10^{11}$
  \msun.  The straight, solid line at the top of the plot shows the
  local stellar mass density as measured by \citet{cole01}.  The shaded region
  illustrates the uncertainty from cosmic variance.}\label{rhoz}
  \plotone{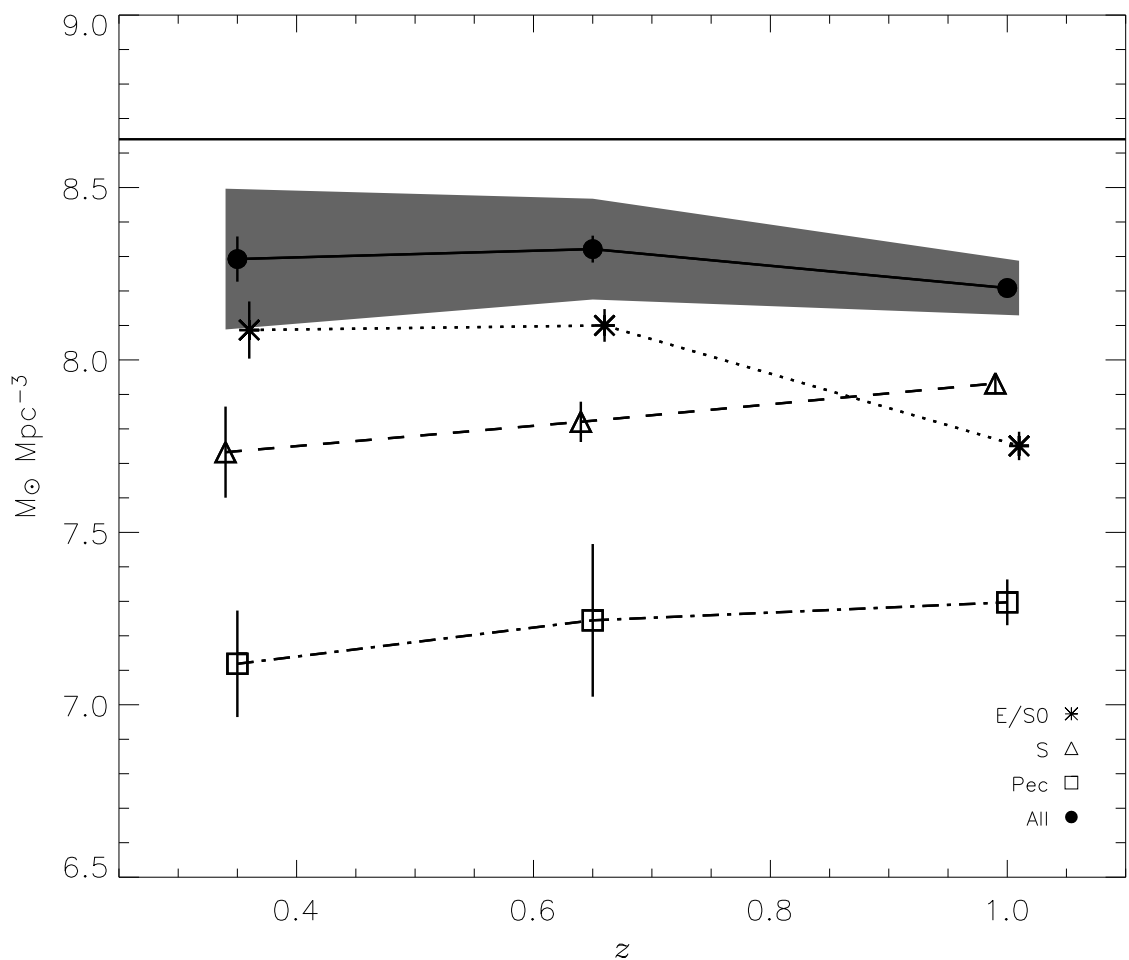}
\end{figure}

As we discuss in $\S$\ref{tot_mfn}, the observed stellar mass function,
which is not affected by incompleteness at high mass in our survey,
shows little evolution from $z \sim 1$ to $z=0$ in agreement with
previous work \citep[e.g.][]{drory04, fontana04, bell04}.  Although it
appears that the stellar content of massive galaxies is relatively well
established by $z \sim 1$, the nature of these galaxies continues to
change.  This effect can be seen in Figure \ref{morph_mfn} in the mass
bin centered at $10^{11}$ \msun~($10.8 < \log M_*/M_{\odot} < 11.2$)
where, in all three redshift intervals, our sample is complete with
respect to a maximum $M_*/L_K$ ratio.  Ellipticals, Spirals, and
Peculiars contribute in similar numbers at the highest redshift, while
at the lowest redshift, Ellipticals clearly dominate.  This trend is
similar to the growth in stellar mass density of these morphological
populations as observed by \citet{BE00} and illustrated in Figure
\ref{rhoz}.

Studying this growth as a function of mass provides additional
information on how Ellipticals come to dominate the massive galaxy
population by $z \sim 0.3$.  This growth of Ellipticals could arise from
several processes including the formation of completely new galaxies,
stellar mass accretion onto established systems of smaller mass, and the
transformation of other established galaxies into ones with early-type
morphologies.  It is likely that all three processes contribute at some
level, although the fact that the density of high-mass galaxies changes
little since $z \sim 1$ places an important constraint on the amount of
new growth, either through star formation or the build-up of smaller
galaxies, that is possible.  This growth is limited to at most a factor
of $\sim$2 while the density of Ellipticals at $M_* \approx 10^{11}$
\msun~increases by a factor of $\sim$3.  A more detailed examination of
the effect of stellar mass accretion onto lower mass Ellipticals is
difficult because of incompleteness in our survey.  We estimate,
however, that the typical stellar mass of an Elliptical with $M_*
\approx 3 \times 10^{10}$ \msun~would have to at least triple in order
to contribute significantly to the increasing proportion of high-mass
Ellipticals.  This spectacular growth in a just a few Gyr seems
unlikely.  Finally, Figure \ref{morph_mfn} shows that the rise of
massive Ellipticals is accompanied by a {\em decline} in the other two
populations.  Thus, assuming that galaxies cannot be broken up or
destroyed, morphological transformation of individual galaxies, whether
through merging or some other mechanism like disk fading, must be a
significant driver in the evolution toward early-type morphologies.

At $z \sim 1$, the higher number of massive Peculiars---often associated
with interacting systems---is suggestive that at least some of this
morphological transformation is occuring through mergers, which are
known to be more frequent at high-$z$ \citep[e.g][]{lef00,
conselice03b}.  Indeed, \citet{hammer04} emphasize the importance of
%
%
luminous IR galaxies (LIRGs), thought to be starbursts resulting from
merging at these masses.  And \citet{bundy04b}, examined galaxies with a
broader range of stellar mass, and used observations of the IR pair
fraction to estimate that the total stellar mass accreted through
merging since $z \sim 1$ is approximately $\Delta \rho_*^m \sim 10^8$
\msun Mpc$^{-3}$.  As illustrated in Figure \ref{rhoz}, this is close to
the magnitude of the observed growth in the mass density of Ellipticals
($\Delta \rho_* \sim 8 \times 10^7$ \msun Mpc$^{-3}$), implying that
merging may have an impact on the morphological transformation
occuring among massive galaxies.

Turning now to the full range of stellar mass accessible in our sample,
one of the key results of this work is the observation of a transitional
mass, $M_{tr}$, above which Ellipticals dominate the mass function and
below which Spirals dominate.  This phenomenon is observed in studies at
low redshift.  \citet{bell03} find a cross-over point at $M_{tr} \approx
3 \times 10^{10}$ \msun~between the local stellar mass functions of
early and late type galaxies, as classified by the concentration index
(see the $g$-band derived stellar mass functions in their Figure 17).
Analysis of the Sloan Digital Sky Survey (SDSS) reveals a similar value
for $M_{tr}$ \citep{kauff03b, baldry04} which also serves as the
dividing line in stellar mass for a number of bimodal galaxy properties
separating early from late types.  These include spectral age
diagnostics (like D$_n$(4000) and H$\delta_A$), surface mass densities,
size and concentration, and the frequency of recent star bursts.

Theoretical work by \citet{dekel04} suggests that, in addition to
mergers, the effects of shock heating and supernova (SN) feedback weigh
heavily on the evolution of galaxies, and both set characteristic scales
that correspond to a stellar mass of $3 \times 10^{10}$ \msun~today.
Galaxies form in halos below the $M_{shock}$ threshold.  If they remain
below the SN feedback threshold, then SN feedback regulates star
formation, leading to young, blue populations and a series of well
defined scaling relations.  Galaxies with masses above $M_{shock}$ are
formed by the merging of progenitors.  Their gas is shock heated and may
be prevented from cooling by AGN feedback, leading to older, redder
populations and a different set of scaling laws
\citep[see][]{birnboim03}.  The SN characteristic scale is expected to
decrease with redshift while the shock heating scale remains constant
(Dekel, private communication), so it is unclear how these physical
scales are related to the evolution of $M_{tr}$ for $z \lesssim 1$.

Whatever the processes at work, the interplay of the morphological
populations and the possible evolution in $M_{tr}$ as traced by Figure
\ref{morph_mfn} seem to echo patterns in the global star formation rate
as observed at higher redshift; the most morphologically evolved
galaxies appear first at the highest masses, and their dominance over
other populations spreads toward lower masses---thereby reducing $M_{tr}$
---as time goes on.  This is similar to the concept
of ``downsizing'' \citep{cowie96} in which the highest mass galaxies
stop forming stars at the earliest times while progressively less
massive galaxies end their star formation later.  Although the
evolution of star formation and morphology both appear to proceed first
at the highest masses, downsizing as it relates to star formation
predates morphological transformation by at least 1--2 Gyr.  As reported
by \citet{juneau04}, the SFR of high-mass galaxies ($M_* > 6 \times
10^{10}$ \msun) goes through a transition at $z \approx 1.5$, emerging from a
``burst phase'' (in which the SFR multiplied by the age of the
universe becomes comparable to the stellar mass) to become
quiescent.  According to Figure \ref{morph_mfn}, these same high-mass
galaxies are still evolving morphologically at $z \approx 0.8$, more than
1 Gyr later.  This implies that the timescale of the transformation
process is $\sim$1 Gyr, similar to the merger timescale, which is often
estimated at 0.5 Gyr \citep[e.g.][]{patton00}.

Downsizing may apply not only to galaxies undergoing bursts of vigorous
star formation on short timescales but also to relatively quiescent
galaxies with continuing, modest star formation.  \citet{treu04}, for
example, find that while massive spheroidals have mostly completed their
star formation by $z \sim 2$, low mass field ellipticals exhibit
continuing star formation from $z < 1$.  Thus, the pattern of downsizing
in both star formation and morphology is gradual and appears to operate
over a large range in mass and extend to the lowest redshifts.

The growing evidence for downsizing and its morphological extension
raises many questions.  While mergers offer a natural explanation for
the link between the continuing star formation and morphological
transformations we present in this paper, it is not clear what is
driving this mass-dependent downsizing behavior.  Most likely several
competing physical processes, including mass-dependent galaxy mergers,
are responsible for shaping the Hubble Sequence.

\section{Summary}\label{conclusions}

We have studied the redshift dependent mass functions for three distinct
morphological populations in a sample of 2150 galaxies with $z_{AB} <
22.5$ in the GOODS fields.  For 44\% of the sample, spectroscopic
redshifts from the KTRS and VLT VIMOS surveys are available.  We use
photometric redshifts from the COMBO17 survey in GOODS-S for 37\% of the
sample and estimate photometric redshifts based on {\it BViz} ACS
photometry for the remaining 19\%.  We utilize \kband~observations of
GOODS-N taken with the WIRC camera at Palomar Observatory and public \kband~imaging
of GOODS-S from the EIS survey to estimate stellar masses for the whole
sample based on fitted mass-to-light ratios.

We find very little evolution in the shape of the combined mass
function, which we fit using Schechter functions with slope, $\alpha
\approx -1.2$ and $\log_{10}M^*h^2 \approx 10.8$---$11.0$ over the whole
redshift range studied ($0.2 < z < 1.4$).  This is consistent with
\citet{fontana04} and appears similar to \citet{drory04} though
\citet{drory04} interpret their results as evidence for stronger
evolution.  Cosmic variance resulting from the small size of the GOODS
fields is our primary source of uncertainty.  The lack of significant
evolution in the observed mass function implies that much of the stellar
growth occurring since $z \sim 1$ takes place at lower masses not yet
accessible to high-$z$ stellar mass surveys.

Our main result is the type-dependent galaxy stellar mass functions over
three redshift intervals spanning the range, $0.2 < z < 1.4$.  The
morphological breakdown of the most massive galaxies ($M_* \approx
10^{11}$ \msun) changes significantly with redshift.  At $z \sim 1$,
Ellipticals, Spirals, and Peculiars are present in similar numbers.  By
a redshift of 0.3, Ellipticals dominate the high-mass population,
suggesting that merging or some other transformation process is active.

At all redshifts in our sample, Spirals and Peculiars dominate at lower
masses while E/S0's become prominent at higher masses.  The observed
transition mass, $M_{tr} = 2$--$3 \times 10^{10}$ \msun, is similar to
that apparent in lower redshift studies.  There is evidence that
$M_{tr}$ was higher at early times, suggesting a morphological extension
of the ``downsizing'' pattern observed in the star formation rate.  Just
as the most massive galaxies emerge from a phase of rapid star formation
at the earliest times, massive galaxies are also the first to evolve
into predominantly early-type morphologies.  This morphological
transformation is completed 1--2 Gyr after the galaxies leave their
bursting phase.

Finally, we derive the integrated stellar mass densities of the three
populations and find similar results as \citet{BE00}.  We find further
evidence for the transformation of Peculiars as well as Spirals into
early-type galaxies as a function of time.  Based on the observed mass
functions, this transformation process appears to be more important at
lower masses ($M_* \lesssim 10^{11}$ \msun) because the most massive
E/S0's are already in place at $z\sim 1$.

In the future it will be possible to extend this kind of study with the
primary aim of reducing statistical uncertainty and the effects of
cosmic variance.  Large galaxy surveys like DEEP2 \citep{davis02} and
COMBO-17 \citep{rix04} are promising in this regard because they contain
tens of thousands of galaxies spread over a wide area, although we note
that stellar mass studies benefit greatly from spectroscopic redshifts.
Extending the combined mass function to lower masses may help reveal the
nature of star formation from $z \sim 1$ to $z=0$.  At the same time,
reducing cosmic variance will allow for more detailed studies on the
type-dependent evolution of the mass function and its relation to
merging and star formation.

\acknowledgments

We wish to thank the referee for very valuable comments, Tommaso Treu
for help developing the morphological classification scheme and for
useful discussions, and Jarle Brinchmann for advice on the stellar mass
estimator.  We also thank Avishai Dekel for helpful discussions.  

Supported by NSF grant AST-0307859 and NASA STScI grant HST-AR-09920.01-A.

%
%
\begin{deluxetable}{lccccccccccc}
\tabletypesize{\scriptsize}
\rotate
\tablewidth{0pt}
\tablecolumns{12}
\tablecaption{Stellar Mass Function Parameters}
\tablehead{
\colhead{}    &  \multicolumn{3}{c}{Total} & \colhead{} & 
\multicolumn{3}{c}{GOODS-N} & \colhead{} & \multicolumn{3}{c}{GOODS-S}  \\ 
\cline{2-4} \cline{6-8} \cline{10-12} \\ 
\colhead{Sample} & \colhead{$\phi^*$} & \colhead{$\alpha$} &
\colhead{$\log_{10} M^* h^2$} & \colhead{} & \colhead{$\phi^*$} & \colhead{$\alpha$} &
\colhead{$\log_{10} M^* h^2$} & \colhead{} & \colhead{$\phi^*$} & \colhead{$\alpha$} &
\colhead{$\log_{10} M^* h^2$} }

\startdata

\cutinhead{$0.2 < z < 0.55$}
All  & $5.4 \pm 0.9$ & $-1.04 \pm 0.18$ & $10.85 \pm 0.10$ &  & $5.8 \pm 1.3$ & $-1.03 \pm 0.21$ & $10.81 \pm 0.10$ &  & $4.9 \pm 0.5
$ & $-1.05 \pm 0.11$ & $10.86 \pm 0.06$ \\ 
All $z_{AB} < 23.5$  & $4.3 \pm 0.7$ & $-1.16 \pm 0.16$ & $10.94 \pm 0.08$ &  & $4.0 \pm 0.5$ & $-1.18 \pm 0.11$ & $10.99 \pm 0.06$
 &  & $4.1 \pm 1.2$ & $-1.16 \pm 0.27$ & $10.92 \pm 0.13$ \\ 
E/S0 & $4.5 \pm 1.5$ & $-0.42 \pm 0.48$ & $10.70 \pm 0.15$ &  & $3.8 \pm 0.5$ & $-0.56 \pm 0.15$ & $10.74 \pm 0.06$ &  & $5.6 \pm 1.1
$ & $-0.16 \pm 0.47$ & $10.52 \pm 0.09$ \\ 
Spiral & $3.4 \pm 0.7$ & $-1.08 \pm 0.25$ & $10.58 \pm 0.11$ &  & $3.1 \pm 0.5$ & $-1.15 \pm 0.16$ & $10.69 \pm 0.09$ &  & $1.1 \pm 
0.1$ & $-1.35 \pm 0.06$ & $10.90 \pm 0.06$ \\ 
Peculiar & $0.1 \pm 0.0$ & $-1.60 \pm 0.12$ & $11.52 \pm 0.24$ &  & $0.1 \pm 0.0$ & $-1.51 \pm 0.06$ & $11.45 \pm 0.16$ &  & $0.3
 \pm 0.0$ & $-1.60 \pm 0.13$ & $10.71 \pm 0.10$ \\ 
\cutinhead{$0.55 < z < 0.8$}
All  & $11.9 \pm 1.5$ & $-0.51 \pm 0.33$ & $10.67 \pm 0.05$ &  & $10.0 \pm 1.0$ & $-0.45 \pm 0.26$ & $10.65 \pm 0.04$ &  & $15.3 \pm 
1.1$ & $-0.48 \pm 0.20$ & $10.70 \pm 0.03$ \\ 
All $z_{AB} < 23.5$  & $8.6 \pm 1.5$ & $-1.09 \pm 0.39$ & $10.83 \pm 0.06$ &  & $6.2 \pm 1.1$ & $-1.15 \pm 0.22$ & $10.86 \pm 0.07$
 &  & $9.6 \pm 1.8$ & $-1.18 \pm 0.30$ & $10.91 \pm 0.06$ \\ 
E/S0 & $6.0 \pm 0.4$ & $0.22 \pm 0.22$ & $10.55 \pm 0.02$ &  & $4.4 \pm 0.5$ & $0.26 \pm 0.45$ & $10.52 \pm 0.05$ &  & $8.0 \pm 0.6$
 & $0.00 \pm 0.22$ & $10.68 \pm 0.03$ \\ 
Spiral & $5.0 \pm 2.3$ & $-0.69 \pm 1.04$ & $10.66 \pm 0.19$ &  & $4.6 \pm 0.7$ & $-0.68 \pm 0.37$ & $10.66 \pm 0.07$ &  & $6.0 \pm 
0.7$ & $-0.57 \pm 0.40$ & $10.62 \pm 0.05$ \\ 
Peculiar & $1.5 \pm 1.3$ & $-1.01 \pm 1.19$ & $10.55 \pm 0.31$ &  & $0.7 \pm 0.2$ & $-0.81 \pm 0.29$ & $11.18 \pm 0.53$ &  & $1.4
 \pm 0.5$ & $-1.03 \pm 0.31$ & $11.03 \pm 0.36$ \\ 
\cutinhead{$0.8 < z < 1.4$}
All  & $5.3 \pm 0.5$ & $-0.50 \pm 0.36$ & $10.80 \pm 0.03$ &  & $6.8 \pm 0.3$ & $-0.02 \pm 0.16$ & $10.72 \pm 0.01$ &  & $3.1 \pm 0.3
$ & $-0.88 \pm 0.33$ & $10.84 \pm 0.03$ \\ 
All $z_{AB} < 23.5$  & $6.2 \pm 5.4$ & $-1.19 \pm 2.66$ & $10.90 \pm 0.28$ &  & $9.1 \pm 1.2$ & $-0.99 \pm 0.30$ & $10.86 \pm 0.02$
 &  & $3.5 \pm 0.8$ & $-1.50 \pm 0.34$ & $10.92 \pm 0.06$ \\ 
E/S0 & $1.5 \pm 0.2$ & $0.11 \pm 0.45$ & $10.73 \pm 0.04$ &  & $1.8 \pm 0.1$ & $0.64 \pm 0.23$ & $10.62 \pm 0.02$ &  & $1.2 \pm 0.3$
 & $-0.61 \pm 0.64$ & $10.83 \pm 0.11$ \\ 
Spiral & $3.1 \pm 0.4$ & $-0.24 \pm 0.46$ & $10.73 \pm 0.04$ &  & $4.0 \pm 0.5$ & $-0.14 \pm 0.42$ & $10.74 \pm 0.05$ &  & $1.1 \pm 
0.3$ & $-0.75 \pm 0.66$ & $11.02 \pm 0.17$ \\ 
Peculiar & $0.7 \pm 0.4$ & $-1.23 \pm 0.72$ & $10.92 \pm 0.21$ &  & $0.6 \pm 0.2$ & $-0.52 \pm 0.76$ & $10.90 \pm 0.14$ &  & $1.4
 \pm 0.4$ & $-0.74 \pm 0.55$ & $10.60 \pm 0.07$ \\ 

\enddata
\label{schfit}
\end{deluxetable}

\end{document}